\documentclass{article}
\usepackage{spconf,amsmath,graphicx}
\usepackage{amsmath,amsfonts,amssymb}
\usepackage{graphicx}
\usepackage{colortbl}
\usepackage[table]{xcolor}
\usepackage{array}
\usepackage{amssymb}
\usepackage{tikz}
\usepackage{calc}
\usepackage[T1]{fontenc}
\usepackage{adjustbox}
\usepackage{graphicx}
\usepackage{subcaption}
\usepackage{float}
\usepackage{graphicx}
\usepackage{subcaption}
\usepackage{multirow}
\usepackage{booktabs} 
\usepackage{hyperref}  
\usepackage{cleveref}  
\usepackage{enumitem}
\usepackage{tikz}
\usepackage{bbm}
\usepackage{amsmath}
\DeclareMathOperator*{\argmax}{arg\,max}

\usepackage{enumitem}
\setlist{nosep, leftmargin=14pt}

\usepackage{mwe} 

\title{An Ensemble-Based Two-Step Framework for Classification of Pap Smear Cell Images}
\name{Theo Di Piazza$^{1, 2}$ \quad Loic Boussel$^{1, 2}$}
\address{$^{1}$UCBL, INSA Lyon, CNRS, Inserm, CREATIS UMR5220, U1294, Villeurbanne, F-69621, France\\
         $^{2}$Department of Radiology, Croix-Rousse Hospital, Hospices Civils de Lyon, Lyon, France}
\begin{document}
%
\maketitle

\textit{Early detection of cervical cancer is crucial for improving patient outcomes and reducing mortality by identifying precancerous lesions as soon as possible. As a result, the use of pap smear screening has significantly increased, leading to a growing demand for automated tools that can assist cytologists managing their rising workload. To address this, the Pap Smear Cell Classification Challenge (PS3C) has been organized in association with ISBI in 2025. This project aims to promote the development of automated tools for pap smear images classification. The analyzed images are grouped into four categories: healthy, unhealthy, both, and rubbish images which are considered as unsuitable for diagnosis. In this work, we propose a two-stage ensemble approach: first, a neural network determines whether an image is rubbish or not. If not, a second neural network classifies the image as containing a healthy cell, an unhealthy cell, or both.}

\section{Introduction}
\label{sec:introduction}
\begin{figure}[h]
        \includegraphics[width=\columnwidth]{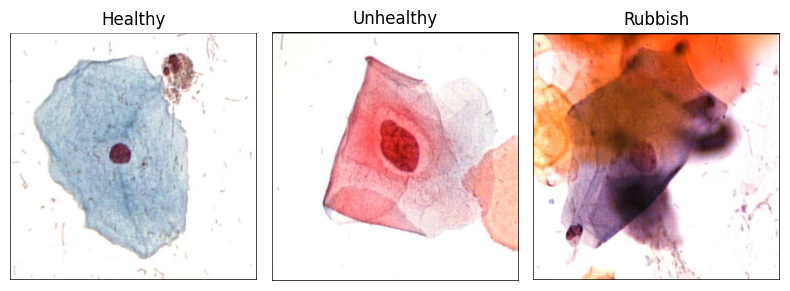}
        \vspace{-2em}
        \caption{\small \textbf{Example of a cell for each class} from the APACC dataset.}
        \label{fig:label_example}
\end{figure}

\begin{figure}[h]
        \includegraphics[width=\columnwidth]{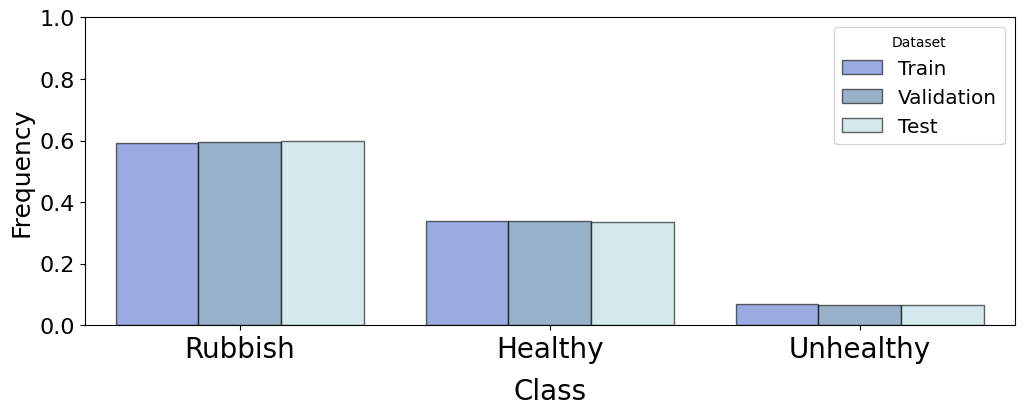}
        \vspace{-2em}
        \caption{\small \textbf{Frequency of classes} in the train, validation and test sets from the APACC dataset.}
        \label{fig:labels_freq}
\end{figure}

Cervical cancer remains one of the most prevalent cancers among women, causing over 300,000 deaths annually~\cite{cohen_cervical_2019}. The widespread adoption of Pap smear screening has significantly improved early detection of cancerous lesions~\cite{sachan_study_2018, perkins_cervical_2023}. This process involves collecting cervical cell samples, preparing them as smears on glass slides, and digitizing them using 3D Scanners. The resulting high-resolution images are then divided into smaller patches~\cite{harangi_pixel-wise_2024}, preserving regions likely to contain diagnostically relevant cells, as illustrated in Figure~\ref{fig:label_example}. Cytologists analyze these slices to identify unhealthy cells, but the vast number of images makes this task time-consuming, resource-intensive, and highly dependent on practitioner expertise~\cite{kupas_annotated_2024}. Deep learning models~\cite{zhou_review_2021} offer a promising avenue to enhance this process and support cytologists to manage their increasing workload by classifying cells as healthy or unhealthy. In medical imaging, extensive efforts have been dedicated to develop deep learning methods for various tasks in cell analysis, including segmentation~\cite{shrestha_efficient_2023,bhattarai_deep-learning-based_2024}, detection~\cite{thomas_review_2017,alahmari_review_2024}, and classification~\cite{shifat-e-rabbi_cell_2020,amitay_cellsighter_2023}. However, Pap smear cell classification remains challenging due to the limited number of publicy available dataset~\cite{jantzen_pap-smear_2005,plissiti_sipakmed_2018,rezende_cric_2021}, the presence of images unsuitable for evaluation (e.g., artifacts, poor resolution) and the class imbalance, where unhealthy cells are significantly outnumbered by healthy ones, as illustrated by Figure~\ref{fig:labels_freq}. To address this, the PS3C Challenge introduced the APACC dataset~\cite{kupas_annotated_2024} to facilitate the development and evaluation of algorithms capable of classifying pap smell images.

\begin{figure*}[t]
        \centering
        \includegraphics[width=0.95\linewidth]{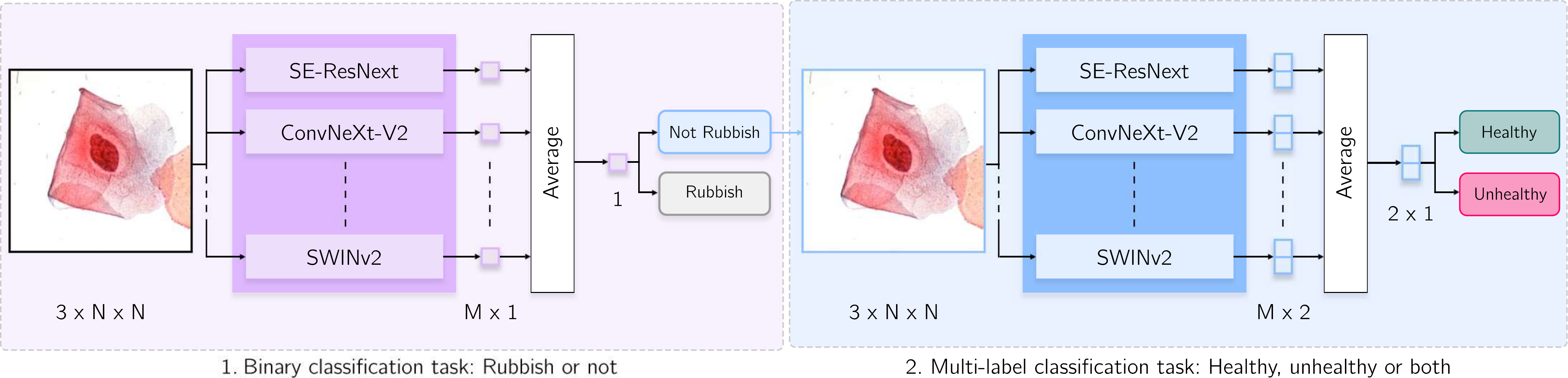}
        \vspace{4pt}
        \caption{Overview of the method. \textbf{Step 1}: Models are independently trained for binary classification to predict whether an image is \textit{rubbish} or not. Final predictions are obtained by averaging the model scores. If the image is classified as non-rubbish, it proceeds to Step 2. \textbf{Step 2}: Models are separately trained for multi-label classification to determine whether the input image contains a \textit{healthy} cell, an \textit{unhealthy} cell, or \textit{both}. Final predictions are computed as the average of model predictions.}
        \label{fig:teaser}
\end{figure*}

Inspired by the diagnostic workflow of cytologists, we propose a two-stage ensemble-based approach. The first stage involves training a model to classify images as either diagnostically suitable or rubbish. In the second stage, a separate model is applied to suitable images to determine the presence of healthy or unhealthy cells, as illustrated by Figure~\ref{fig:teaser}.

Our contributions can be summarized as follows:
\begin{itemize}[noitemsep,topsep=0pt,parsep=0pt,partopsep=0pt]
    \item A two-step framework leveraging ensemble learning to boost classification performance on Pap Smear Cell data.
    \item A benchmarking of state-of-the-art methods on the APACC public dataset, providing a robust comparison framework.
    \item We release the source code at \url{https://github.com/theodpzz/ps3c}.
\end{itemize} 

\section{Related work}
\label{sec:related_work}

\subsection{Convolutional Neural Network} Early approaches to visual recognition~\cite{wu_efficient_2022} leveraged convolutional neural networks to extract feature representations from images. Due to their ability to capture hierarchical features through local receptive fields, CNNs have been widely adopted across various applications, including industrial inspection~\cite{staar_anomaly_2019}, medical imaging~\cite{zhou_review_2021}, and remote sensing~\cite{zhu_deep_2017}. Deep convolutional networks have demonstrated robust performance across a range of tasks, including classification~\cite{wu_deep_2024}, segmentation~\cite{minaee_image_2020}, and detection~\cite{zhao_object_2019}. However, their training becomes increasingly challenging as model complexity grows. ResNet introduced residual connections between layers of varying depths, facilitating training and improving performance~\cite{he_deep_2015}. More recently, the introduction of Squeeze-and-Excitation (SE) Networks~\cite{hu_squeeze-and-excitation_2019} led to the development of SE-ResNeXt, an enhancement of ResNeXt~\cite{xie_aggregated_2017} that improves performance in visual recognition. Inspired by the development of Vision Transformers~\cite{dosovitskiy_image_2021}, ConvNeXt~\cite{liu_convnet_2022} was proposed as an evolution of traditional convolutional networks, incorporating architectural elements from Vision Transformers, such as layer normalization~\cite{ba_layer_2016} and improved regularization~\cite{huang_densely_2018}.

\subsection{Vision Transformer} 

The attention mechanism~\cite{vaswani_attention_2023}, originally introduced in Natural Language Processing, has shown strong performance across various text-related tasks~\cite{devlin_bert_2018,radford_language_nodate,touvron_llama_2023}. This mechanism was quickly adapted to the vision domain with the introduction of Vision Transformers~\cite{dosovitskiy_image_2021}, which model images as a set of fixed-size patches that interact through the attention mechanism, enabling a better global context understanding compared to CNNs, which struggle to capture long-range dependencies. More recently, the Swin Transformer~\cite{liu_swin_2021,liu_swin_2022} leveraged hierarchical shifted windows to constrain attention computation to local neighborhoods of image patches, improving both local and global context understanding, and achieving superior performance across several vision tasks~\cite{pereira_review_2024}.

\subsection{Ensemble Deep Learning} Ensemble methods combine multiple models within a unified framework to enhance performance~\cite{mohammed_comprehensive_2023,polikar_ensemble_2012}. Rather than relying on a single model, these methods aggregate the predictions from several models to leverage the strengths of diverse architectures while mitigating the weaknesses of individual approaches. Common ensemble techniques include bagging~\cite{breiman_bagging_1996}, boosting~\cite{freund_experiments_nodate}, and stacking~\cite{ganaie_ensemble_2022}. Bagging trains base models independently and aggregates their predictions, reducing variance and overfitting~\cite{breiman_random_2001}. Boosting, on the other hand, trains models sequentially, where each iteration focuses on correcting the errors of the previous model to reduce bias~\cite{freund_efcient_nodate}. Stacking involves training models separately and then using a meta-learner—typically a small, independent neural network—to learn how to optimally combine the base model predictions, further improving performance~\cite{hospedales_meta-learning_2020}.

\section{Method}
\label{sec:method}

As illustrated in Figure~\ref{fig:teaser}, we employ a two-stage approach. First, an initial model determines whether the image is classified as rubbish or not. If the image is not considered as rubbish, it is then processed by a second model, which predicts whether a healthy or unhealthy cell is observed.

\subsection{Dataset preparation}
\label{section:dataset}

We use the APACC public dataset~\cite{kupas_annotated_2024} to train and evaluate our method. APACC contains 103,675 cervical cell images extracted from 107 smears (for 107 unique patients) and 4 distinct types of classes (\textit{rubbish}, \textit{healthy}, \textit{unhealthy} and \textit{both}) annotated by domain experts. We employ a 5-fold cross-validation strategy~\cite{kohavi_study_nodate}. For each fold, the dataset is divided into training, validation, and test sets following an 80-10-10 split ratio, ensuring a balanced distribution of classes across all subsets, as illustrated by Figure~\ref{fig:labels_freq}. We use images labeled \textit{both} exclusively in the training sets, as they do not appear in the final test set. Since our ensemble method leverages multiple models, images are resized according to the input resolution of each model (see Table~\ref{tab:complexity_comparison}). To ensure compatibility with pretrained networks, we normalize images using ImageNet dataset statistics~\cite{noauthor_image_nodate}. For data augmentation, we apply the following transformations: horizontal and vertical flips, resized crop, elastic transform, and rotation.

\subsection{Step 1: Binary classification task}
\label{section:binar_classif}
The first step of our framework involves predicting whether an image should be classified as rubbish or not, formulated as a binary classification task. During training, we consider two labels: \textit{rubbish} and \textit{non-rubbish} (where \textit{healthy}, \textit{unhealthy}, and \textit{both} are grouped under the \textit{non-rubbish} label). Each image $x \in \mathbb{R}^{3 \times N \times N}$ is processed by a backbone network, denoted as $\Phi_{1}: \mathbb{R}^{3 \times N \times N} \to \mathbb{R}^{d}$, pre-trained on ImageNet~\cite{noauthor_imagenet_nodate}, to extract a feature representation $h \in \mathbb{R}^{d}$, such that:
\begin{equation}
    h_{1} = \Phi_{1}(x) \, ,
\end{equation}

The representation $h$ is then passed through a classification head, denoted as $\Psi_{1}: \mathbb{R}^{d} \to \mathbb{R}$ implemented as a lightweight multilayer perceptron, which produces a logit score $\hat{y_{1}} \in \mathbb{R}$:
\begin{equation}
    \hat{y_{1}} = (\Psi_{1} \circ \Phi_{1})(x) = \Psi_{1}(h_{1}) \, .
\end{equation}

We experiment with $M \in \mathbb{N}^{+}$ backbones, including Vision Transformers~\cite{dosovitskiy_image_2021}, SWINv2~\cite{liu_swin_2022}, ConvNeXt-V2~\cite{liu_convnet_2022} and SE-ResNeXt~\cite{xie_aggregated_2017}. This diverse selection encompasses both convolution-based and attention-based architectures, trained on varying input resolutions. The model is optimized for binary classification using the Binary Cross-Entropy loss. A Sigmoid is applied to turn logits into probabilities.

\subsection{Step 2: Multi-label classification task}
\label{section:multilabel_classif}
The second stage of our framework aims to predict the presence of \textit{healthy} or \textit{unhealthy} cells within a given input image. Since both labels can be present simultaneously in the same image, we formulate this as a multi-label classification task with two target labels: \textit{healthy} and \textit{unhealthy}. For training, we only consider \textit{non-rubbish} images from the training, validation, and test sets. Similar to Stage 1, the input image $x$ is first processed by a backbone network, noted as $\Phi_{2} : \mathbb{R}^{3 \times N \times N} \to \mathbb{R}^{d}$, that extracts a feature representation $h_{2} \in \mathbb{R}^{d}$. This representation $h_{2}$ is then passed through a classification head noted $\Psi_{2} : \mathbb{R}^{d} \to \mathbb{R}^{2}$, producing a logit vector $\hat{y_{2}}$ where each component corresponds to a prediction score for the respective label (\textit{healthy} or \textit{unhealthy}), formulated as:
\begin{equation}
    \hat{y_{2}} = (\Psi_{2} \circ \Phi_{2})(x) = \Psi_{1}(h_{2}) \, .
\end{equation}

The model is trained using a multi-label classification objective with a Cross-Entropy loss function, with class weights balancing class frequencies. A Sigmoid is applied to turn logits into probabilities.

\subsection{Ensemble method}
\label{section:ensemble_method}
\noindent \textbf{Step 1.} The $M \in \mathbb{N}^{+}$ models are trained independently on each fold $j \in \{1, \ldots, 5\}$. Given an input image $x$, each model $i \in \{1, \ldots, M\}$ outputs a probability $p^{\text{rubbish}}_{i, j} \in [0, 1]$. The final prediction probability, denoted as $p^{\text{rubbish}}_{j} \in [0, 1]$ is obtained by averaging the individual model probabilities, as followed:
\begin{equation}
    p_{j}^{\text{rubbish}} = \frac{1}{M} \sum_{i=1}^{M} p^{\text{rubbish}}_{i, j} \, .
\end{equation}

Since we employ a 5-fold cross-validation strategy, this results in 5 probability, denoted $\{p^{\text{rubbish}}_{1}, \ldots, p^{\text{rubbish}}_{5}\}$. For each fold $j$, we compute the threshold that maximizes the F1-score on the validation set and apply it to the probabilities of the test set to get the corresponding prediction $c_{1, j}$. The final prediction $c_{1} \in \{\text{rubbish}, \text{suitable}\}$ is obtained via majority voting, where the label most frequently predicted across the 5 folds is selected.
\begin{equation}
    c_{1} = \argmax_{c \in \{\text{rubbish}, \text{suitable}\}} \sum_{j=1}^{5} \mathbbm{1}(c_{1, j} = c) \, .
\end{equation}

\noindent \textbf{Step 2.} If an image is not classified as \textit{rubbish} in Step 1, it proceeds to Step 2. Similar to Step 1, we derive a probability for each fold $j$ for the \textit{healthy} label noted as $p_{j}^{\text{healthy}} \in [0, 1]$ by averaging predictions from the $M$ models across the fold. For each fold $j$, we then select the threshold $t_{j} \in [0, 1]$ that maximizes the macro-F1 score on the validation set and apply it to the predictions of the test set, as follows:
    \begin{equation}
        c_{2, j} =
\begin{cases}
\textit{healthy}, & \text{if } p_{j}^{\text{healthy}} \geq t_{j} \\
\textit{unhealthy}, & \text{otherwise.}
\end{cases} \,
    \end{equation}

The final prediction $c_2 \in \{\text{healthy}, \text{unhealthy}\}$ is obtained through majority voting, corresponding to the most frequent label predicted across the 5 folds, as follows:
\begin{equation}
    c_{2} = \argmax_{c \in \{\text{healthy}, \text{unhealthy}\}} \sum_{j=1}^{5} \mathbbm{1}(c_{2, j} = c) \, .
\end{equation}

\begin{table*}[h]
\centering
\begin{tabular}{l c c c c c c}
\toprule%
Method & F1 Score & Weighted F1 Score & Precision & Recall & AUROC & Accuracy\\
\toprule
\small Random Prediction & \small $29.22 \text{\scriptsize $\pm 0.29$}$ & \small $26.76 \text{\scriptsize $\pm 0.33$}$ & \small $33.38 \text{\scriptsize $\pm 0.34$}$ & \small $33.29 \text{\scriptsize $\pm 0.23$}$ & \small $49.92 \text{\scriptsize $\pm 0.15$}$ & \small $33.32 \text{\scriptsize $\pm 0.36$}$\\
\textbf{ViT-L} & \small $75.74 \text{\scriptsize $\pm 0.80$}$ & \small $88.06 \text{\scriptsize $\pm 0.33$}$ & \small $79.99 \text{\scriptsize $\pm 1.26$}$ & \small $73.14 \text{\scriptsize $\pm 1.17$}$ & \small $84.95 \text{\scriptsize $\pm 0.64$}$ & \small $82.20 \text{\scriptsize $\pm 0.20$}$\\
\textbf{SwinV2-B} & \small $75.80 \text{\scriptsize $\pm 1.05$}$ & \small $88.22 \text{\scriptsize $\pm 0.35$}$ & \small $80.71 \text{\scriptsize $\pm 1.07$}$ & \small $72.91 \text{\scriptsize $\pm 1.16$}$ & \small $85.39 \text{\scriptsize $\pm 0.22$}$ & \small $92.32 \text{\scriptsize $\pm 0.29$}$\\
\textbf{SwinV2-L} & \small $76.12 \text{\scriptsize $\pm 1.16$}$ & \small $87.79 \text{\scriptsize $\pm 0.50$}$ & \small $80.84 \text{\scriptsize $\pm 1.11$}$ & \small $73.27 \text{\scriptsize $\pm 1.98$}$ & \small $85.40 \text{\scriptsize $\pm 0.56$}$ & \small $92.05 \text{\scriptsize $\pm 0.27$}$\\
\textbf{SE-ResNeXt} & \small $76.65 \text{\scriptsize $\pm 1.49$}$ & \small $88.22 \text{\scriptsize $\pm 0.39$}$ & \small $\mathbf{81.13} \text{\scriptsize $\pm 1.40$}$ & \small $73.83 \text{\scriptsize $\pm 1.66$}$ & \small $85.20 \text{\scriptsize $\pm 0.67$}$ & \small $92.31 \text{\scriptsize $\pm 0.28$}$\\
\textbf{ConvNeXt-V2} & \small $\underline{76.92} \text{\scriptsize $\pm 1.32$}$ & \small $\underline{88.42} \text{\scriptsize $\pm 0.31$}$ & \small $\underline{80.98} \text{\scriptsize $\pm 1.84$}$ & \small $\underline{74.49} \text{\scriptsize $\pm 2.20$}$ & \small $\underline{85.68} \text{\scriptsize $\pm 1.16$}$ & \small $\underline{92.41} \text{\scriptsize $\pm 0.18$}$\\
\rowcolor[gray]{0.93} 
\textbf{Ensemble learning} & \small $\mathbf{78.46} \text{\scriptsize $\pm 1.17$}$ & \small $\mathbf{89.08} \text{\scriptsize $\pm 0.38$}$ & \small $80.94 \text{\scriptsize $\pm 1.61$}$ & \small $\mathbf{76.63} \text{\scriptsize $\pm 1.67$}$ & \small $\mathbf{86.98} \text{\scriptsize $\pm 0.41$}$ & \small $\mathbf{92.82} \text{\scriptsize $\pm 0.26$}$\\
\toprule
\end{tabular}
\caption{\textbf{Quantitative evaluation on the APACC test set.} Reported mean and standard deviation metrics were computed over a 5-fold Cross-Validation. \textbf{Best} results are in bold, \underline{second best} are underlined.}\label{tab:quantitative}
\end{table*}

\section{Experimental setup}
\label{sec:experimental_setup}

For the first and second steps, each model was trained with a batch size of $32$, using the AdamW~\cite{loshchilov_decoupled_2019} optimizer for 80 epochs, with a learning rate of $10^{-5}$. The training required a GPU with 80GB of memory.
 
\section{Results}
\label{sec:results}

\subsection{Quantitative results}

We evaluate the performance of our approach using standard (macro) classification metrics: AUROC, Accuracy, Precision (P), Recall (R), and F1-Score, the latter being the harmonic mean of Precision and Recall~\cite{rainio_evaluation_2024}. We also report the weighted F1-Score, computed as the label-frequency-weighted average of per-class F1-Scores on the test set. The reported values represent the mean scores across 5-fold cross-validation. Table~\ref{tab:quantitative} reports the classification metrics for each model in our ensemble approach. All configurations are trained and evaluated using a two-step prediction process (Figure~\ref{fig:teaser}). Vision Transformer (ViT-L)~\cite{dosovitskiy_image_2021} achieves a macro F1-score of $75.75$ and an AUROC of $84.95$. SwinV2-Large~\cite{liu_swin_2022} improves performance with an F1-score of $76.12$ ($\Delta$+0.49\% over ViT-L). SE-ResNeXt, leveraging Squeeze-and-Excitation blocks~\cite{hu_squeeze-and-excitation_2019}, further enhances results with a $\Delta$+0.67\% increase in F1-score compared to SwinV2-Large. ConvNeXt-V2~\cite{woo_convnext_2023} achieves the highest individual performance, reaching an F1-score of $76.92$ ($\Delta$+1.56\% over ViT, $\Delta$+1.05\% over SwinV2-Large, and $\Delta$+0.35\% over SE-ResNeXt). Our ensemble method, averaging model probabilities, achieves an F1-score of $78.46$, surpassing ConvNeXt-V2 by $\Delta$+2.00\%. As shown in Table~\ref{tab:complexity_comparison}, ConvNeXt-V2 is trained with a $384 \times 384$ resolution and a higher latent space dimensionality than other backbones, suggesting increased expressiveness in its learned representations. Table~\ref{tab:per_class_f1} reports the per-class F1-score for the categories \textit{rubbish}, \textit{healthy}, and \textit{unhealthy}. The relative ranking of methods remains consistent across all metrics, aligning with the macro-F1 trends.

\begin{table}[h]
    \centering
    \begin{tabular}{l c c c c}
        \hline
        \small Method & \small Resolution & \small Emb. dim. & \small FLOPs & \small F1\\ \hline
        \small ViT-L  & \small $384^2$ & \small $1024$ & \small $349$ & \small $75.75 \text{\scriptsize $\pm 0.80$}$\\
        \small SwinV2-B   & \small $256^2$ & \small $1024$ & \small $40$ & \small $75.80 \text{\scriptsize $\pm 1.05$}$\\
        \small SwinV2-L   & \small $384^2$ & \small $1536$ & \small $403$ & \small $76.12 \text{\scriptsize $\pm 1.16$}$\\
        \small SE-ResNeXt   & \small $288^2$ & \small $2048$ & \small $57$ & \small $76.65 \text{\scriptsize $\pm 1.48$}$\\
        \small ConvNeXt-V2  & \small $384^2$  & \small $2816$ & \small $675$ & \small $76.92 \text{\scriptsize $\pm 1.32$}$\\
        \hline
    \end{tabular}
    \caption{\textbf{Comparison of different backbones on APACC classification}. The Emb. dim. column corresponds to the dimension of the feature extracted by the corresponding backbone. FLOPs column refers to the number of floating-point operations (in giga, G).}
    \label{tab:complexity_comparison}
\end{table}

\begin{table}[h]
    \centering
    \begin{tabular}{l c c c c}
        \hline
        \small Method & \small Rubbish & \small Healthy & \small Unhealthy\\ \hline
        \small ViT-L   & $91.76 \text{\scriptsize $\pm 0.15$}$ & $84.75 \text{\scriptsize $\pm 0.70$}$ & $50.74 \text{\scriptsize $\pm 2.38$}$\\
        \small SwinV2-B   & $91.87 \text{\scriptsize $\pm 0.29$}$ & $85.02 \text{\scriptsize $\pm 0.56$}$ & $50.51 \text{\scriptsize $\pm 2.56$}$\\
        \small SwinV2-L   & $91.56 \text{\scriptsize $\pm 0.35$}$ & $84.16 \text{\scriptsize $\pm 1.11$}$ & $52.64 \text{\scriptsize $\pm 3.69$}$\\
        \small SE-ResNeXt   & $91.79 \text{\scriptsize $\pm 0.29$}$ & $84.92 \text{\scriptsize $\pm 0.36$}$ & $53.22 \text{\scriptsize $\pm 4.08$}$\\
        \small ConvNeXt-V2   & $\underline{91.96} \text{\scriptsize $\pm 0.15$}$ & $\underline{85.17} \text{\scriptsize $\pm 0.35$}$ & $\underline{53.63} \text{\scriptsize $\pm 3.57$}$\\
        \small Ensemble   & $\mathbf{92.36} \text{\scriptsize $\pm 0.45$}$ & $\mathbf{86.06} \text{\scriptsize $\pm 0.45$}$ & $\mathbf{56.96} \text{\scriptsize $\pm 3.12$}$\\
        \hline
    \end{tabular}
    \caption{\textbf{Per-class F1-Score on the APACC test set.}}
    \label{tab:per_class_f1}
\end{table}

\section{Conclusion}
\label{sec:conclusion}

In this work, we introduced an ensemble-based method to address the challenging task of pap smear cell classification for cervical cancer diagnosis. This problem is particularly difficult due to the presence of non-suitable images for diagnostic and the underrepresentation of \textit{unhealthy} labels. Specifically, we proposed an ensemble of convolutional and transformer-based networks, pretrained on natural images and fine-tuned on the public APACC dataset. Our final model achieved a macro-F1 score of $86.61$ on the final competition test set. Future work could explore alternative ensemble strategies, such as boosting, or incorporate a meta-learner to optimally combine model predictions based on individual performance.


\section{Acknowledgments}
\label{sec:acknowlegments}
We acknowledge Kupas et al. 2024~\cite{kupas_annotated_2024} for making the APACC dataset available. This work was performed using HPC resources from GENCI–[IDRIS] [Grant No. 103718]. We thank the support team of Jean Zay for their assistance.

\bibliographystyle{IEEEbib}
\bibliography{biblio}
\end{document}